\documentclass[twocolumn]{aastex61}
\shorttitle{Calculating \ika Line Emissivities}
\shortauthors{Krawczynski \& Beheshtipour}
\usepackage{amsmath}
\newcommand{\ika}{Fe K$\alpha$ }
\begin{document}
\accepted{for publication in ApJ, 9/18/2017}

\title{On the Calculation of the \ika Line Emissivity of Black Hole Accretion Disks}
\correspondingauthor{Henric Krawczynski}
\email{krawcz@wustl.edu}
\author{H. Krawczynski}
\author{B. Beheshtipour}
\affil{Physics Department and McDonnell Center for the Space Sciences, Washington University in St. Louis,
1 Brookings Drive, CB 1105, St. Louis, MO 63130, USA}

\begin{abstract}
Observations of the fluorescent \ika emission line from the inner accretion flows of stellar mass black holes in 
X-ray binaries and supermassive black holes in Active Galactic Nuclei have become an important tool to study 
the magnitude and inclination of the black hole spin, and the structure of the accretion flow close to the 
event horizon of the black hole. Modeling spectral, timing, and soon also X-ray polarimetric observations 
of the \ika emission requires to calculate the specific intensity in the rest frame of the emitting plasma.
We revisit the derivation of the equation used for calculating the illumination of the accretion disk 
by the corona. We present an alternative derivation leading to a simpler equation, 
and discuss the relation to the previously published results.
\end{abstract}
\section{Introduction \label{intro}}
The \ika line in the X-ray energy spectra of stellar mass black holes in X-ray binaries and 
supermassive black holes in Active Galactic Nuclei (AGN) is thought to originate as 
fluorescent emission when the disk is illuminated by the hard X-rays from a hot corona
\citep[see][for recent reviews]{Mill:15,Reyn:14}. The shape and origin of the corona are still a matter of debate 
although spectral, timing, and -- in the case of gravitationally lensed quasars -- 
the amplitude of microlensing amplification favor extremely compact coronas 
close to the event horizons of the black holes 
\citep[e.g.][]{Reis:13}.  X-ray energy spectra are commonly 
modeled by folding the specific emissivity $I$ with Cunningham's transfer function 
$f$ \citep{Cunn:75,Spei:95}. For a geometrically thin equatorial accretion disk 
of a black hole described by a stationary axis symmetric, and asymptotically flat spacetime 
the luminosity at energy $E$ per unit energy and steradian is given by:     
 \begin{equation} 
\frac{d^2L_{E}}{dE\,d\Omega}\!=\!\! \int_{r_1}^\infty \!\!\!\! \int_0^1\!\!\!\!  
\frac{\pi \,r\,g^2}{\sqrt{g^*(1-g^*)}}
I(E/g,r, \mu)
f(g^*,r,\vartheta_{\rm o})\,
dg^* dr
\end{equation} 
where $r_{1}$ is the inner edge of the accretion disk, $r$ is the radial coordinate,
$g$ is the ratio between the observed and emitted energy of the photon,
$g^*$ is the same ratio rescaled so that the minimum and maximum values for the emission from
the considered accretion disk ring are 0 and 1, respectively,
$\mu$ is the cosine of the polar angle of the emitted emission with respect to the disk normal 
in the rest frame of the accretion disk plasma,and $\vartheta_{\rm o}$ is the inclination of the observer 
measured from the rotation axes of the black hole and the accretion disk (assumed to have co-aligned spin axes).
The $g$ and $g^*$ terms in the integral reduce the $g$-dependence of $f$.
The parameter $g^*$ is double-valued for each ring, and the integral has to be 
performed twice. 

Assuming axial symmetry, the emitted intensity $I$ of the \ika emission follows from convolving 
the hard X-ray flux impinging on the accretion disk $F_{\rm X}$ with a reflection function $R$:
\begin{equation} 
I(E,r,\mu)=\int\int dE'\,d\mu'\,
F_{\rm X}(E',r,\mu')\,R(E,\mu;E',r,\mu').
\label{Chan}
\end{equation} 
Here, $F_{\rm X}(E',r,\mu')$ is the energy flux of the X-ray radiation per unit energy 
at energy $E'$ incident at radial coordinate $r$ from a direction with direction cosine $\mu'$ per unit time and area
(all quantities in the rest frame of the accreting plasma) averaged over the azimuthal incident angles. 
$R$ \citep[defined here in loose analogy 
to Chandrasekhar's scattering function $S$,][Chapter 13]{Chan:60} is the specific intensity of the 
reflected emission with energy $E$ emitted into the directions with direction cosine $\mu$.
We have chosen the most general form of $R$ consistent with Cunningham's transfer function approach, 
and the $r$ dependence could arise for example from the disk ionization changing with $r$.  
The double integral in Equation (\ref{Chan}) is commonly replaced by the product of 
an incident flux times an emission coefficient. A notable exception is \citet{Garc:14} 
who account for the $\mu$-dependence of the reprocessed emission. 
If the corona emission follows a power law distribution 
with a photon index $\Gamma$ ($dN/dE\propto E^{-\Gamma}$) or equivalently, with 
a spectral index $\alpha$ ($E \,dN/dE\propto E^{-\alpha}$ with $\alpha=\Gamma-1$),
the relevant flux is the photon flux $N(>E_{\rm thr})$ above the threshold energy 
$E_{\rm thr}$ required for the emission of an \ika fluorescent photon. 
The integral photon flux above $E_{\rm thr}$ scales with $E_{\rm thr}^{\,-\alpha}$.

As we acquire more and more accurate \ika spectral and timing observations, 
there is a growing interest in pinning down the physical properties (shape, location, energy spectra) 
of the corona, by comparing detailed predictions for specific corona geometries with observations.
For example, \citet{Fuku:07} study the disk illumination by isotropically and an-isotropically emitting 
point-like lamppost coronae. \citet{Wilk:12} analyze lamppost coronae on the symmetry axis as well 
laterally offset coronae orbiting the symmetry axis, disk-shaped coronae, and lamppost corona 
moving along the symmetry axis. \citet{Daus:13} discuss the illumination pattern created by static 
point-like and radially elongated static and accelerating coronas \citep[see also][]{Gonz:17}. 

All of the these studies start with 
a determination of the flux $F_{\rm X}$ impinging on the accretion disk. 
In Section \ref{equations} of this paper, we review the standard 
argument of how to convert the radial distribution of the corona photons 
(derived from integrating the geodesic equations) into the photon flux per proper time and proper area. 
We present a much simpler derivation, obtain a much simpler equation, 
and clarify the relation between our results and the previous results.
We close with a brief summary and outlook in Section \ref{discussion}.
Although we limit the following discussion to General Relativity's Kerr metric in Boyer-Lindquist \citep[BL][]{Boye:67} 
coordinates, the results can easily be adapted to work for other stationary, axisymmetric, and 
asymptotically flat black hole spacetimes \citep[see][for recent reviews describing observational 
constraints on alternative black hole spacetimes]{Joha:16,Bamb:17}. 

We use geometric units ($G$=$c$=1) and express all distances
in units of the gravitational radius $r_{\rm g}=GM/c^2$ with $M$ being the 
mass of the black hole. $a\in\left[-1,1\right]$ denotes the black hole angular 
momentum per unit mass.
\section{The coronal photon flux impinging on the accretion disk per proper time and proper area}
\label{equations}
In terms of the BL coordinates $x^{\mu}=(t,r,\theta,\phi)$, the Kerr metric is given by:
\begin{equation}
ds^2\,=\,g_{00}\,dt^2+2g_{03}\,dt\,d\phi+g_{11}\,dr^2+g_{22}d\theta^2+g_{33}d\phi^2
\end{equation}
with
$g_{00}=-(1-2r/\Sigma)$, 
$g_{03}=-2ar{\rm\,sin}^2\theta/\Sigma$,
$g_{11}=\Sigma/\Delta$, 
$g_{22}=\Sigma$, and
$g_{33}=(r^2+a^2+2a^2r{\rm\,sin}^2\theta/\Sigma){\rm\,sin}^2\theta$,
$\Sigma=r^2+a^2{\rm\,cos}^2\theta$, and $\Delta=r^2-2r+a^2$.

\begin{figure}
\plotone{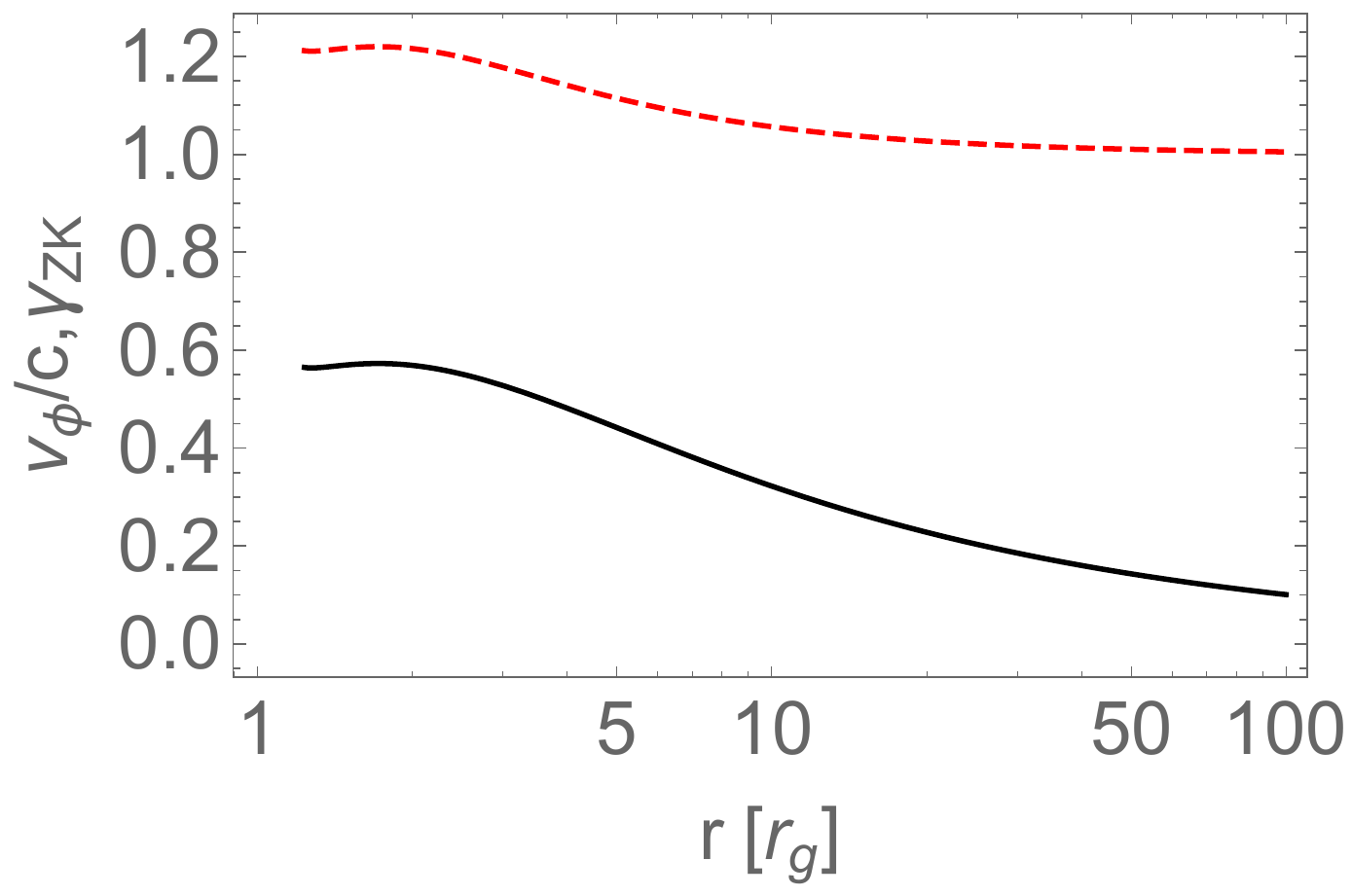}
\caption{\label{gamma} Velocity $v_{\phi}$ in units of the speed of light (solid lower line) 
and Lorentz factor $\gamma_{\rm ZK}$ (dashed upper line) 
of the accretion disk plasma (assumed to orbit the black hole on a Keplerian orbit) as measured by a 
ZAMO for a black hole spin of $a=0.998$ commonly used as the maximum spin 
of an astrophysical black hole \citep{Thor:74}.}
\end{figure} 
We introduce two observers, the {Zero Angular Momentum Observer} (ZAMO, subscript Z) 
and the {Keplerian observer} (KO, subscript K) orbiting the black hole with the angular 
frequencies $\omega=-g_{03}/g_{33}=2 a/(r^3+a^2 r+2 a^2)$ and $\Omega=\pm(r^{3/2}\pm a)^{-1}$, respectively.
The upper sign refers to direct orbits, and the lower to retrograde orbits. In the following we limit
the discussion to direct orbits. As in standard thin disk theory, we assume that the accretion 
disk plasma orbits the black hole on Keplerian orbits, and the KO frame is thus the rest frame of the accretion disk plasma. 
The four velocities of the two observers are given 
by ${\bf u}_{\rm Z}$ and ${\bf u}_{\rm K}$ with the contravariant components \citep{Bard:72}:
\begin{eqnarray}
u_{\rm Z}^{\,\mu}&=&u_{\rm Z}^{\,0}(1,0,0,\omega)\\
u_{\rm K}^{\,\mu}&=&u_{\rm K}^{\,0}(1,0,0,\Omega).
\label{Kepler}
\end{eqnarray}
Normalizing the four velocities to -1 gives the zero components 
$u_{\rm Z}^{\,0}= \sqrt{(r^2+a^2+2a^2/r)/(r^2-2r+a^2)}$ and 
$u_{\rm K}^{\,0}=(a+r^{3/2})/\sqrt{r^3-3r^2+2ar^{3/2}}$. 
 
For an observer moving with four velocity ${\bf u}$ we associate a tetrad 
\citep[a systemof orthogonal and normalized basis vectors of the tangent vector space][Chapter 7]{Chan:83} 
by defining a time-like basis vector  ${\bf e}_{(0)}={\bf u}$ and three space-like basis vectors
${\bf e}_{(1)}=A \partial_1$,
${\bf e}_{(2)}=B \partial_2$, and
${\bf e}_{(3)}=C \partial_{0}+D \partial_3$.
The orthonormality conditions ${\bf e}_{(i)}\cdot{\bf e}_{(j)}=\delta^i_{\, j}$ ($i,j=1...3$)
together with ${\bf e}_{(0)}\cdot{\bf e}_{(i)}=0$ determine the values of $A$, $B$, $C$ and $D$
(see the appendix for the ZAMO and KO values).
Given the tetrad for a point ${\cal P}$ and the coordinates 
of an event in an infinitesimal neighborhood of ${\cal P}$ in tetrad coordinates $x^{\mu'}$, 
the BL coordinates $x^{\nu}$ are given by:
\begin{equation}
x^{\nu}(x^{\mu'})\,=\,x^{\nu}({\cal P})+x^{\mu'}e_{(\mu')}^{\nu}
\end{equation}
where we sum over $\mu'$. 
The relation gives the transformation matrix for 
tangent vectors $\Lambda^{\nu}_{\,\,\mu'}=e_{(\mu')}^{\nu}$ and
we obtain the inverse transformation by matrix inversion.
\begin{figure}
\plotone{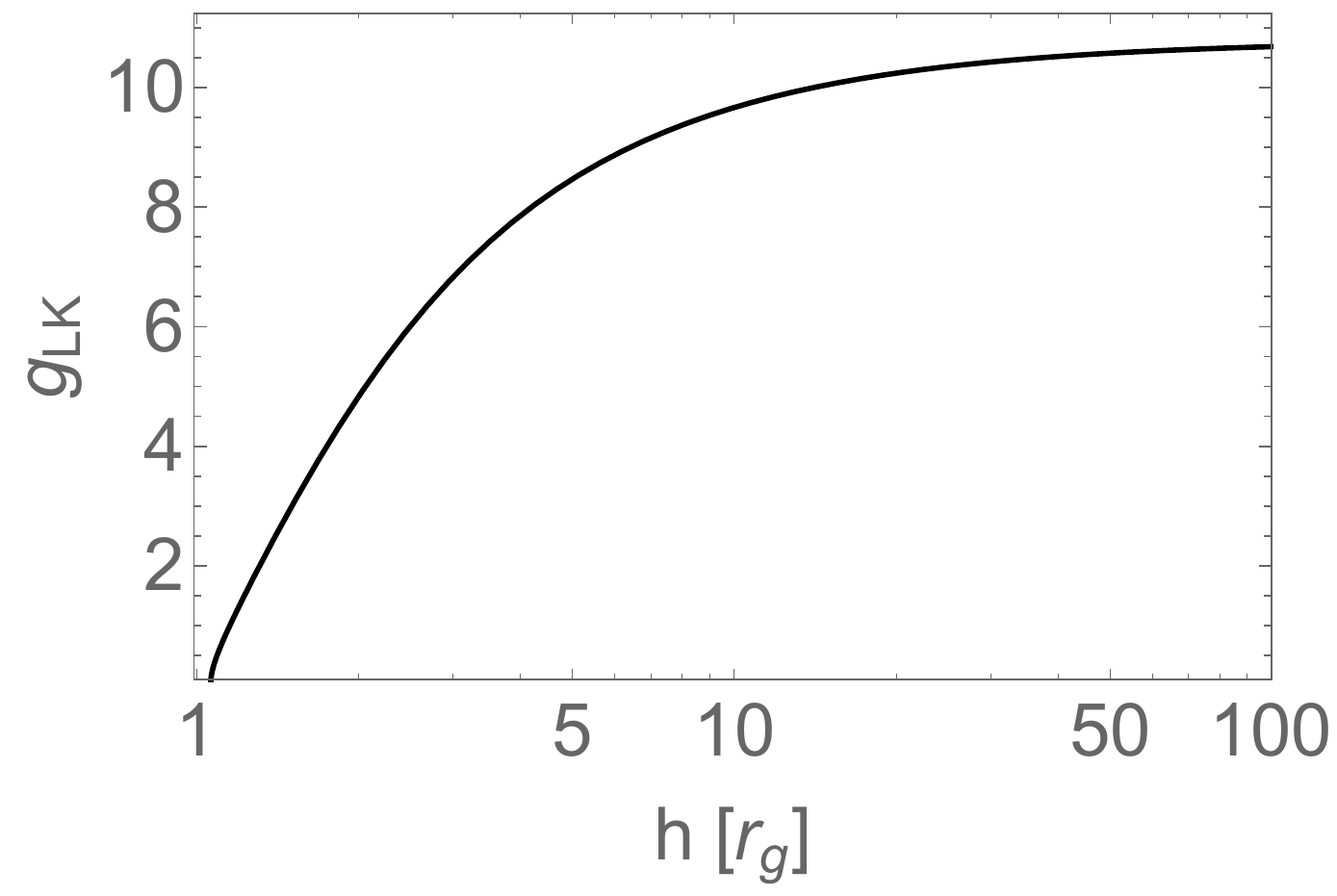}
\caption{\label{glk} Assuming a black hole with spin $a=0.998$, the graph shows the
ratio $g_{\rm LK}$ of the frequencies between emission and absorption of a photon 
traveling from the lamppost corona at height $h$ to the inner edge of the accretion disk
at $r_{\rm ISCO}(a)=1.24\,r_{\rm g}$. The ratio is smaller than 1 for $h<1.06$.
}
\end{figure} 

\subsection{Traditional Derivation}
In the following we refer to the photon flux hitting the accretion disk per proper time and proper area as $f$ and the 
corresponding energy flux as $F$.  The standard way of determining $f$ or $F$ consists of integrating 
the geodesics of photons leaving the corona to determine $N(r,dr)$, the number of photon trajectories intersecting 
the equatorial plane with a radial coordinate between $r$ and $r+dr$ per considered time interval, 
and calculating $f$ with an equation of the form 
\citep[e.g.][]{Wilk:12,Daus:13,Gonz:17}:
\begin{equation}  
f(r)=\,\frac{g_{\rm LK}}{\gamma_{\rm ZK} A_{\rm Z}(r,dr)} N(r,dr)
\label{wf12}
\end{equation}  
Here, $A_{\rm Z}(r,dr)$ is the proper area of an accretion disk ring extending from $r$ to $r+dr$ 
as measured by a ZAMO:
\begin{eqnarray}
A_{\rm Z}(r,dr)&=&
\left(
\int_0^{2\pi} \left|\frac{\partial (x^{1'},x^{3'})}{\partial (r,\phi)}\right|
d\phi
\right)
dr\nonumber \\
&=&\!2\pi \sqrt{g_{11}g_{33}}\,dr\!\!=\!\!2\pi \sqrt{\frac{r^4+a^2 r^2+2 a^2 r}{r^2-2 r+a^2}}\!dr
\end{eqnarray}  
The factor $\gamma_{\rm ZK}$ is the Lorentz factor of the KO relative to the ZAMO, and describes 
by how much larger the ring area is for the (faster moving) KO than for the ZAMO.
We obtain the $\phi$-component of the relative velocity and the corresponding Lorentz factor with the equations:
\begin{eqnarray}
v_{\phi}&=&\frac{{\bf u}_{\rm K}\cdot {\bf e}_{(3)}}{-{\bf u}_{\rm K}\cdot {\bf e}_{(0)}}\nonumber \\
&=&\frac{r^{5/2}-2 a r+a^2 \sqrt{r}}{\left(r^{3/2}+a\right) \sqrt{{r^3-2r^2+a^2 r}} }
\\
\gamma_{\rm ZK}&=&\frac{1}{\sqrt{1-v_{\phi}^{\,2}}}.
\end{eqnarray}
Figure \ref{gamma} shows $v_{\phi}$ and $\gamma_{\rm ZK}$ for a rapidly spinning black hole.

Finally, $g_{\rm LK}$ is the relative change of the photon energy between its arrival at the accretion disk 
(measured by a co-rotating KO) and its emission (measured in the reference frame of the lamppost corona).
The factor transforms the photon emission rate in the lamppost reference frame 
to the emission rate observed by a KO.   
The ratio can be calculated by projecting the photon's four velocity onto the four velocity of the 
disk plasma and lamppost corona, respectively:
\begin{equation}
g_{\rm LK}=\frac{{\bf u}_{K}\cdot {\bf p}_{\gamma,K}}
{{\bf u}_{L}\cdot {\bf p}_{\gamma,L}}.
\end{equation}
where 
${\bf u}_{K}$ is the four velocity of the KO from Equation (\ref{Kepler}), and 
\begin{equation}
{\bf u}_{\rm L}^{\,\mu}=((g_{00})_{\rm L}^{\,-1/2},0,0,0)
\end{equation}
is the four velocity of the lamppost corona with the subscript L denoting the evaluation at the lamppost position.
The four momentum of a null geodesic of the Kerr spacetime can be parameterized 
as follows \citep[][Sect. 62, Eq. 187]{Chan:83}:
\begin{equation}
p_{\rm K, \mu}=(E,\sqrt{{\cal R}}/\Delta,\sqrt{\Theta},-L_{\rm z})
\end{equation}
with $E$ being the photon energy, 
${\cal R}$ and $\Theta$ being functions which depend on the coordinates and the constants of motion,
and $L_{\rm z}$ being the photon's angular momentum with respect to the symmetry axis.
As $L_{\rm z}$ vanishes for photons from a lamppost corona, $g_{\rm LK}$ is given by:
\begin{equation}
g_{\rm LK}=\frac{u_{\rm K}^{\,\,0}}{u_{\rm L}^{\,\,0}}.
\end{equation}
Figure \ref{glk} shows $g_{\rm LK}$ for a rapidly spinning black hole.

Combining the results and simplifying considerably, we obtain:
\begin{equation}
f(r)=\frac{1}{2\pi r}\sqrt{1-\frac{2h}{a^2+h^2}}\frac{dN(r,dr)}{dr}
\end{equation}
The energy flux per proper time and proper area is obtained by multiplying $f(r)$ with the photon energy
in the lamppost frame $E_{\rm L}$ times $g_{\rm LK}$:
\begin{equation}
F(r)=g_{\rm LK}\,E_{\rm L}\,f(r)
\end{equation}

\subsection{Alternative Derivation}
\begin{figure}
\begin{center}
\includegraphics[scale=.4]{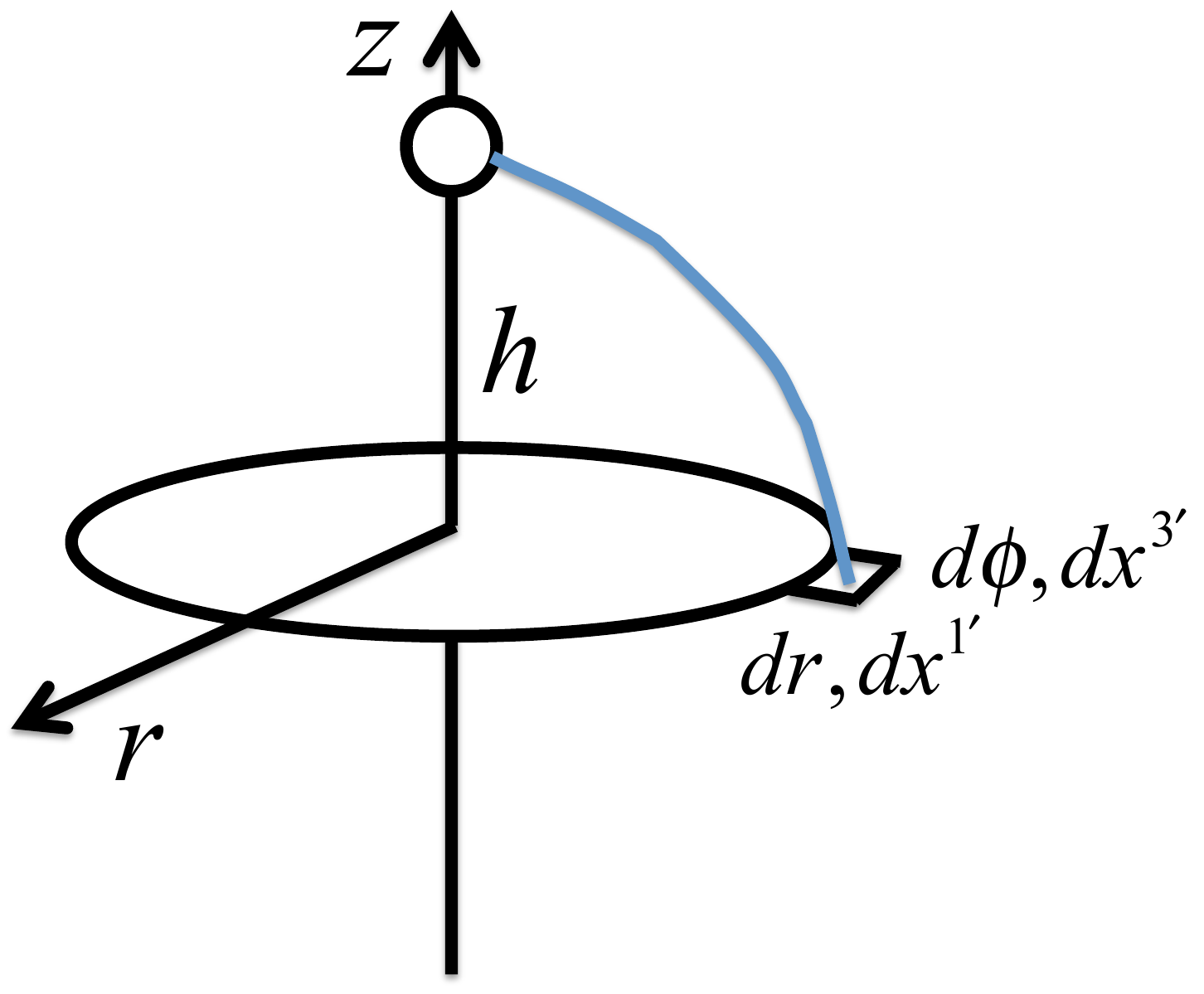}
\end{center}
\caption{\label{sketch01} Sketch of the lamppost corona illuminating an infinitesimal area element of the accretion disk.}
\end{figure}

We refer to Figure \ref{sketch01} for an alternative derivation. We consider photons emanating  from the 
lamppost corona for a time $dt$ in BL coordinate time, and consider that the accretion disk ring from $r$ to $r+dr$ 
is made of infinitesimally small area elements. For each area element the ``exposure'' ${\cal E}$ of the area element 
to corona photons equals the product of the proper time $dx^{0'}$ and the proper area $dx^{1'} dx^{2'}$.
The dashed coordinates refer to the coordinates of a KO co-rotating with the accretion disk plasma. 

We find the exposure of all area elements of the ring by integrating the locally defined exposure ${\cal E}$ over 
the global BL coordinate $\phi$ using the Jacobian for the transformation from 
$(x^{0'}, x^{1'}, x^{3'})$ to $(t,r\phi)$ to express $dx^{0'}dx^{1'}x^{3'}$ in terms of $dt$, $dr$, and $d\phi$:
\[
{\cal E}=\left(\int_0^{2\pi} 
\left|
\frac{\partial (x^{0'}, x^{1'}, x^{3'})}{\partial (t, r, \phi)}
\right| d\phi\right)dr\, dt
\]
\begin{equation}
= \left(\int_0^{2\pi}
\sqrt{-g_{\rm tr\phi}} d\phi\right) dr\, dt=2\pi r \,dr \,dt.
\end{equation}
The second line follows from the general result that the Jacobian is given by the square root of the 
ratio of the metric determinants  $\sqrt{g_{tr\phi}/g'_{013}}$ 
with $g_{tr\phi}=-r^2$ being the determinant of the $t,r$, and $\phi$ part of the metric in BL coordinates, 
and $g'_{013}=-1$ being the determinant of the 0, 1, and 3 part of the metric in the KO coordinates 
\citep[e.g.][Section 1.7]{Pois:04}.   

\begin{figure}
\plotone{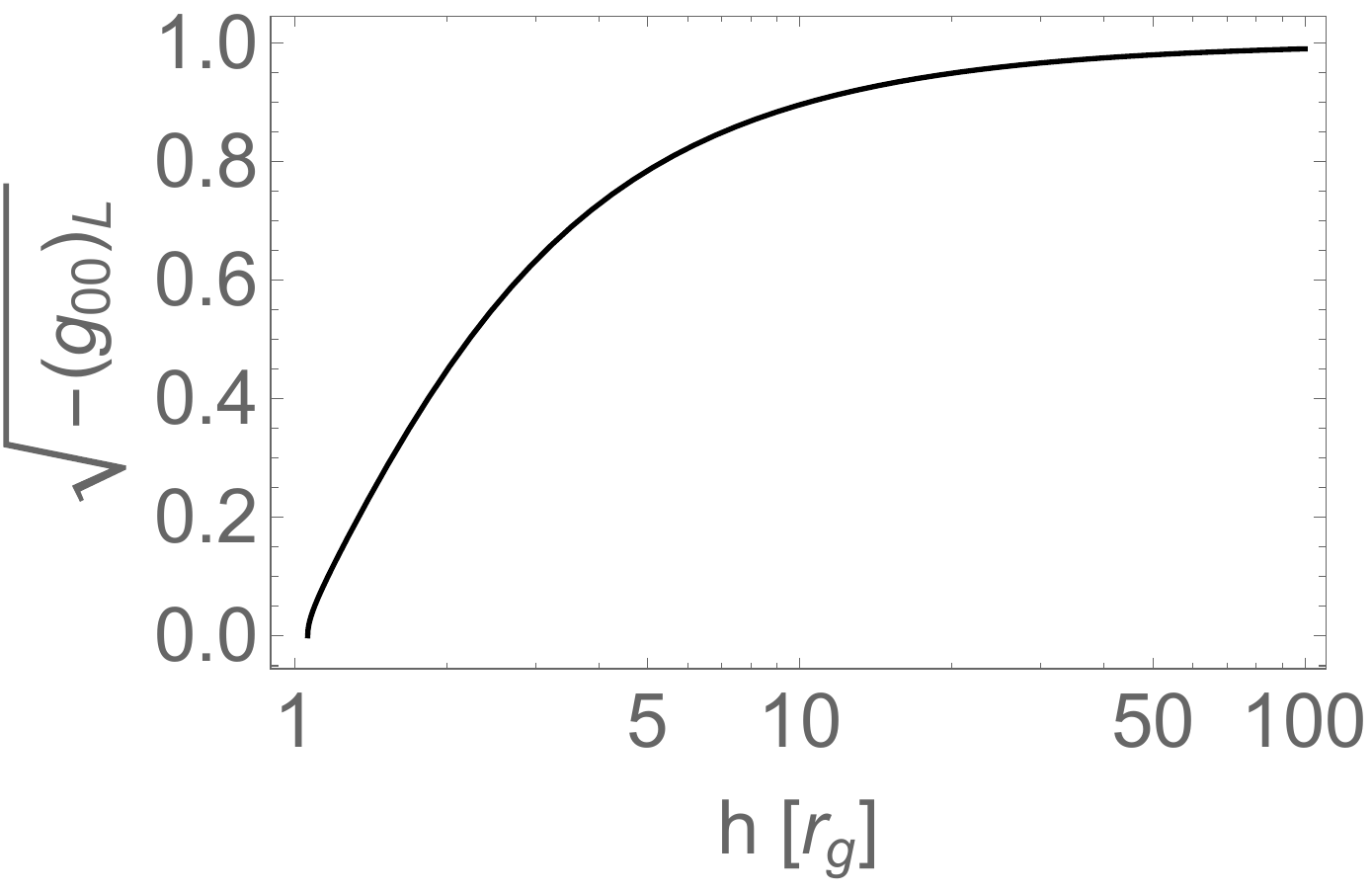}
\caption{\label{g00} Time dilation factor between the coordinate time $t$ and the lamppost proper time.
}
\end{figure}
Dividing the number of photons $N(r,dr)$ impinging on the accretion disk ring 
from $r$ to $r+dr$ in the coordinate time interval $dt$ by the exposure, 
we obtain the photon flux per proper time and proper area:    
\begin{equation}
f_2(r)=\,\frac{1}{2\pi r}\frac{dN(r,dr)}{dr}
\label{simple}
\end{equation}
Equation \ref{simple} is considerably simpler than Equation (\ref{wf12}). 
Somewhat amusingly, discussing Equation (\ref{wf12}) \citet{Wilk:12}  
noted that ``the general relativistic effects on the area of orbiting annuli 
in the accretion disc are cancelled exactly by one factor of the redshift''.
Indeed, the $g_{\rm LK}$ in the numerator of Equation (\ref{wf12}) 
approximately cancels the $\gamma_{\rm ZK}$ in the denominator.
This cannot be an exact cancellation as $g_{\rm LK}$ depends on the height
of the lamppost while $\gamma_{\rm ZK}$ does not.

What is the relation between Equations (\ref{wf12}) and (\ref{simple})?
The attentive reader will anticipate the answer: whereas the latter equation converts the number of photons 
$N(r,dr)$ impinging on the ring {\em per unit coordinate time} into the photon flux per unit proper time and unit proper area, 
the former converts the number of photons $N(r,dr)$ impinging on the ring {\em per unit lamppost time} 
into the photon flux per unit proper time and unit proper area. 
Indeed, converting $f_2$ from ``per coordinate time'' into ``per lamppost time'' we recover $f$:
\begin{eqnarray}
f_2(r)\frac{dt}{dt_{\rm L}}&=&f_2(r) \frac{1}{u_{\rm L}^{\,0}}
=f_2(r)\sqrt{-(g_{00})_{\rm L}}\nonumber\\
&=&f_2(r)\sqrt{1-\frac{2h}{a^2+h^2}}=f(r)
\label{simple2}
\end{eqnarray}
The proportionality factor $\sqrt{-(g_{00})_{\rm L}}$ is shown in Figure \ref{g00}.
$f$ is lower than $f_2$ as a certain photon emission rate in the lamppost frame 
corresponds to a lower rate measured in Boyer-Lindquist coordinate time.
\section{Discussion}
\label{discussion}
In this paper, we discuss equations for determining the photon flux per proper time 
and proper area impinging on the accretion disk of a black hole in the Kerr metric. 
Our discussion clarifies the assumptions underlying the derivation of the standard equations 
used in the literature. In particular, the quantity $N(r,dr)$ in Equation (\ref{wf12}) refers to the 
rate of photons impinging on an accretion disk ring extending from $r$ to $r+dr$ per unit lamppost time. 
We present a conceptually much simpler derivation which does not require the transformation into the 
auxiliary ZAMO reference frame and which relies on the concept of a relativistically invariant
proper exposure (product of exposure time and area). 

In the near term, {\it Chandra}, {\it XMM-Newton}, {\it Swift}, and {\it NuSTAR} 
will continue to give new spectroscopy and timing \ika data. The results derived here can be 
used for calculating the \ika emissivity via Cunningham's transfer function. 
Around 2020, NASA's and ESA's {\it Imaging X-ray Polarimetry Explorer (IXPE)} \citep{Weiss:16} will add 
qualitatively new information. Interestingly, this will require modelers to update their existing 
codes, as Cunningham's transfer function does not account for the change of the 
polarization direction along the photon geodesics.

Our current work focuses on deriving the emissivity profiles for a number of 3-D corona models, 
and to compare the predictions with a large body of observational timing and spectral 
results (Beheshtipour et al., in preparation).
\section*{Appendix}
The constants appearing in the definition of the ZAMO tetrad are given by:
\begin{eqnarray*}
A&=&g_{11}^{\,-1/2},\,\,\,
B=g_{22}^{\,-1/2}\\
C&=&0,\,\,\,D=(r^2+a^2+2 a^2/r)^{\,-1/2}
\end{eqnarray*}
For the Keplerian observer the constants read:
\begin{eqnarray*}
A&=&g_{11}^{\,-1/2},\,\,\,
B=g_{22}^{\,-1/2}\\
C&=&{\frac{r^2-2 a \sqrt{r}+a^2}{\sqrt{(r^3-3r^2+2a r^{3/2}) \left(r^2-2r+a^2\right)}}}\\ 
D&=&\frac{r^{3/2}-2\sqrt{r}+a}{\sqrt{(r^3-3r^2+2a r^{3/2})(r^2-2r+a^2)}}
\end{eqnarray*}
\section*{Acknowledgments}
We thank NASA (grant \#NNX14AD19G) for financial support, and the 
anonymous referee for excellent comments. 

\end{document}